\documentclass[conference]{IEEEtran}
\IEEEoverridecommandlockouts
\usepackage[numbers,sort&compress]{natbib}
\usepackage{amsmath,amssymb,amsfonts}
\usepackage{algorithmic}
\usepackage{graphicx}
\usepackage{textcomp}
\usepackage{xcolor}
\usepackage[ruled]{algorithm2e}
\usepackage{multirow} 
\usepackage{pifont}
\def\BibTeX{{\rm B\kern-.05em{\sc i\kern-.025em b}\kern-.08em
    T\kern-.1667em\lower.7ex\hbox{E}\kern-.125emX}}
\begin{document}

\title{Image Compression and Reconstruction Based on Quantum Network\\

}
\author{
    \IEEEauthorblockN{Xun Ji$^{1,2,3}$, Qin Liu$^{1,3}$, Shan Huang$^{1,2}$, Andi Chen$^{1,3}$, Shengjun Wu$^{1,2,3,*}$}
    \IEEEauthorblockA{$^1$National Laboratory of Solid State Microstructure, School of Physics, \\and Collaborative Innovation Center of Advanced Microstructures, Nanjing University, Nanjing 210093, China}
    \IEEEauthorblockA{$^2$Institute for Brain Sciences and Kuang Yaming Honors School, Nanjing University, Nanjing 210023, China}
    \IEEEauthorblockA{$^3$Hefei National Laboratory, University of Science and Technology of China, Hefei 230088, China}
    \IEEEauthorblockA{Email: \{mg20980002, 5022022220137, hs, 502022980001\}@smail.nju.edu.cn, sjwu@nju.edu.cn}
}

\maketitle

\begin{abstract}
Quantum network is an emerging type of network structure that leverages the principles of quantum mechanics to transmit and process information. Compared with classical data reconstruction algorithms, quantum networks make image reconstruction more efficient and accurate. They can also process more complex image information using fewer bits and faster parallel computing capabilities. Therefore, this paper will discuss image reconstruction methods based on our quantum network and explore their potential applications in image processing. We will introduce the basic structure of the quantum network, the process of image compression and reconstruction, and the specific parameter training method. Through this study, we can achieve a classical image reconstruction accuracy of 97.57\%. Our quantum network design will introduce novel ideas and methods for image reconstruction in the future. \\
\end{abstract}

\begin{IEEEkeywords}
image compression and reconstruction, quantum network, parameter training
\end{IEEEkeywords}

\section{Introduction}
With the wide application of digital images in various fields, signal compression and reconstruction is a research hotspot in digital signal and image processing \cite{b1,b2}. 
The compression and reconstruction algorithm of the classical process encodes classical data into sparse vectors by classical sparse coding (CSC) algorithm and then reconstructs the original input by a linear combination of sparse coding vectors and dictionary sets \cite{b3,b4}. The CSC algorithm can be used in image reconstruction, sparse image coding, signal coding, pattern recognition, and image enhancement. After encoding and compressing the image data, the image file size can be effectively reduced, thus saving storage space and transmission bandwidth \cite{b5}. In image reconstruction, the classical algorithm can restore the original image by decoding and decompressing \cite{b6}, ensuring image quality \cite{b7}. In recent years, with the development of deep learning and artificial intelligence technology, image compression and reconstruction algorithms have been further improved \cite{b8}, bringing new development opportunities to the field of digital image processing \cite{b9,b10}. 

Quantum network and quantum algorithm, as a new computing mode based on the principle of quantum mechanics, have also attracted much attention in image processing \cite{b11,b12,b13,b14,b15}. Quantum computing has great potential in image processing because of its parallel computing ability and the properties of quantum states \cite{b12,b13}. Quantum networks can realize efficient image compression and reconstruction, which are expected to process large-scale image data and bring new possibilities for image processing technology \cite{b11,b14}. The application of quantum compression algorithms in image processing has also been widely discussed, such as data compression based on quantum PCA algorithm \cite{b11}, quantum image compression and reconstruction algorithm \cite{b12,b13,b14}, and a series of research results have been obtained.  Compression and reconstruction of signals by quantum computing can significantly reduce problems in large data flows during data transmission \cite{b13}. Therefore, based on classical compression and reconstruction algorithms, we propose a quantum compression reconstruction algorithm to solve the classical image reconstruction problem \cite{b14,b15}. Of course, the algorithm will solve quantum problems more intelligently.

In this paper, an image compression and reconstruction algorithm based on quantum network is designed by combining classical image compression and reconstruction algorithm \cite{b5} with the quantum algorithm \cite{b16}. First, the classical image information is encoded into quantum states (quantum information) \cite{b17,b18}. Then the prepared quantum states are input into the quantum compression and reconstruction network, which are achieved by optical circuits \cite{b19,b20,b21}. 
In the process, we measure the output state of the quantum compression network and the output state of the quantum reconstruction network, respectively. Then the measurements are converted into the compressed image and the reconstructed image, which are utilized to train the parameters of the quantum network based on the gradient descent algorithm. Finally, the simulation of compression and reconstruction of grayscale images is effectively realized by this quantum algorithm. This algorithm's application and development trend in digital image processing is discussed \cite{b22}. The fusion of image processing and quantum computing has promoted the development of digital image processing technology. Next, we will  explain quantum compression and reconstruction algorithms in detail.

\begin{figure*}[htbp]\label{F1}
\centerline{\includegraphics[scale=0.75]{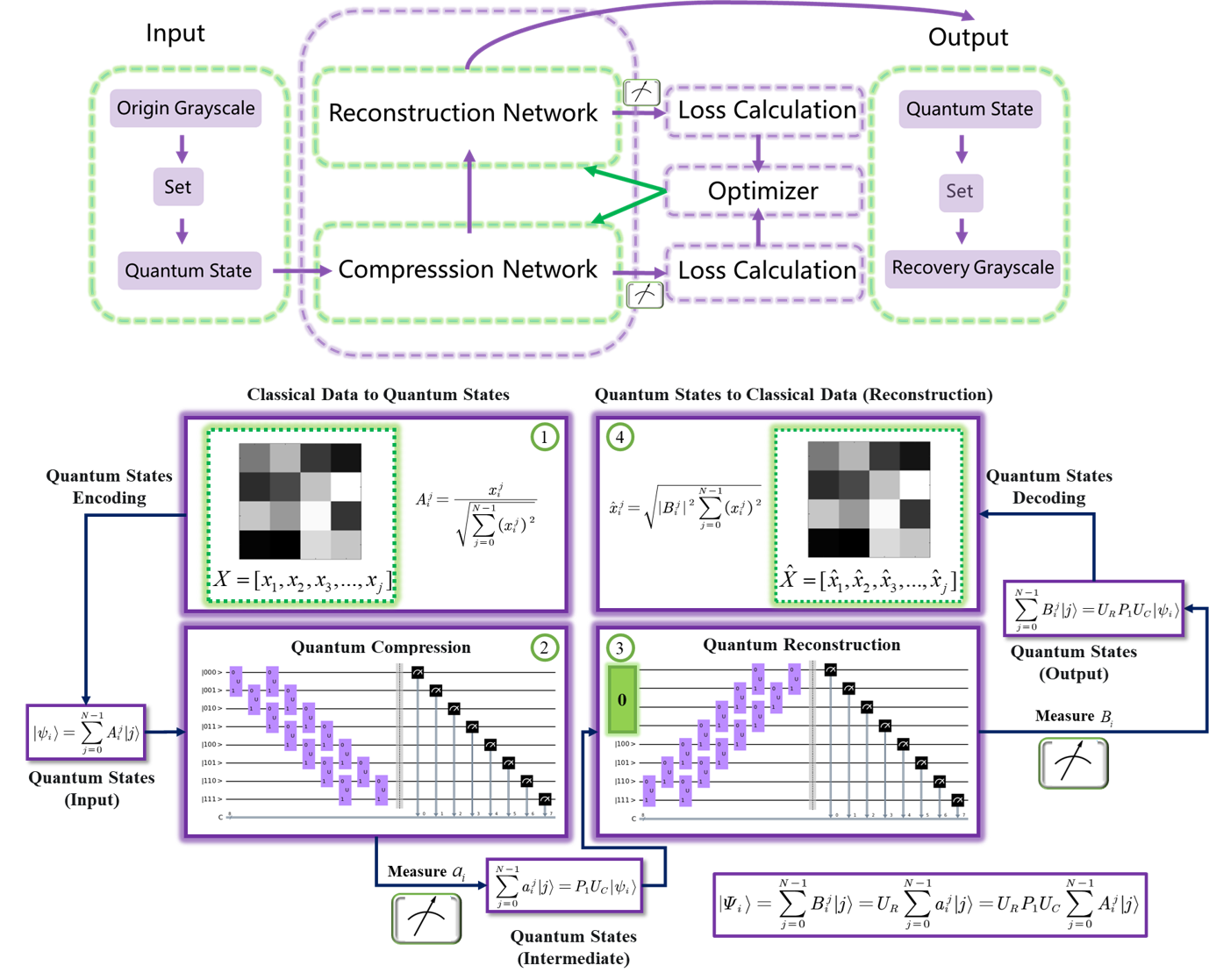}}
\caption{Image Compression and Reconstruction Based on Quantum Network. \ding{172} Encode the classical image data $x_i$ into the probability amplitudes $A_i^j$ of quantum states $|\psi _i\rangle$. \ding{173} Input the prepared quantum states $|\psi _i\rangle$ into the compression network $U_C$ to obtain the quantum states in lower $d$-dimensional space. In this process, the output states in $U_C$ are measured, the parameter gradient is calculated based on the loss function, and the optimal parameters in $U_C$ are fed back by the optimizer. \ding{174} Input the quantum states in the $d$-dimensional space into the reconstruction network $U_R$ to obtain the output states $|\Psi _i\rangle$ in the higher-dimensional space. The parameter training process in $U_R$ is the same as in $U_C$. \ding{175} The probability $B_i^j$ of output states in the reconstruction network is converted into classical image data $\hat{x}_i$ to realize image reconstruction.}
\end{figure*}

\section{preliminary}
\subsection{Encode Images to Quantum States}
For image or general classical data, the data matrix can be converted into a $N$-dimensional row vector $X$. According to the basic principles of quantum states, we encode the elements in classical data vector $X$ as the probability amplitudes $A^j$ of the quantum state $|\psi \rangle=\sum{A}^j|j\rangle$. For $N$-dimensional data, at least $\lceil \text{log}_2(N)\rceil$ qubits are required; for example, if the data is in 16 dimensions ($N=16$), four qubits are needed to implement 16-dimensional data encoding ($n=4$). The specific coding rules are as follows:
\begin{align}\label{eq1}
		A_i^j=\frac{x_i^j}{\sqrt{\sum_{j=0}^{N-1}\left(x_i^j\right)^2}},
\end{align}
 where $i$  $\in [1,2,3,...,M]$, and $M$ is the number of image samples. The dimension of data $j$ satisfies that $j\in[0,1,2,..,N-1]$. $x_i^j$ represents the $j$-th dimensional data in the $i$-th image samples. $A_i^j$ represents the  probability amplitude of  the $j$-th computational basis state in the $i$-th quantum states \cite{b16}. Therefore, the $i$-th quantum states are concluded as $|\psi _i\rangle=\sum_{j=0}^{N-1}A_i^j|j\rangle$.

On the contrary, the measurements of output states can also be converted to classical data. As follows,  $\hat{x}$ represents the reconstruction of classical information, and we need to retain the sum of squares in the input data $\left(x_i^j\right)^2$ to decompile states to data. The states are decoded into data, which can be described as: 
\begin{align}\label{eq2}
		\hat{x}_i^j=\sqrt{\left|B_i^j\right|^2 \sum_{j=0}^{N-1}\left(x_i^j\right)^2} ,
\end{align}
where classical data $\hat{x}_i^j$ is obtained from measurement of the final reconstructed output states $|\Psi_i\rangle=\sum_{j=0}^{N-1}B_i^j|j\rangle$, and the input data is effectively compressed and reconstructed. Both compression and reconstruction networks reach the target optimal solution, and the data reconstruction $\hat{x}_i^j$ is almost the same as the original information data $x_i^j$. And $\hat{X}$ is obtained by combining all reconstructed classical output data. 

\subsection{Image Compression Based on Quantum Network}
Image compression and reconstruction based on quantum network are divided into two independent networks, which can be represented as a quantum compression network $U_C$ and a quantum reconstruction network $U_R$ respectively, as shown in Fig. 1. In the former, prepared quantum states $|\psi _i\rangle$ are input into the compressed network $U_C$, and the output of $U_c$ can be represented as follows:
\begin{align}\label{eq3}
    |\Phi_i\rangle=\sum_{j=0}^{N-1}{a_{i}^{j}}|j\rangle =P_1U_C|\psi _i\rangle=P_1U_C\sum_{j=0}^{N-1}A_i^j|j\rangle,
\end{align}
where $a_{i}^{j}$ is the probability amplitude of the $j$-th computational basis state in the $i$-th output state in $U_C$ and $P_1$ is the projection transformation by which $d$-dimensional states can be compressed as states $|\Phi_i\rangle$. 
The identity matrix ($I$) can consist of $P_1$ and $P_0$, namely, $P_1+P_0=I$, as shown in Fig. 2. By adjusting $P_1$ and $P_0$, we can achieve compression with different space sizes. Furthermore, by measuring the output states of the compression network, the loss error can be calculated according to the given image compression target to optimize the network parameters and reduce the loss.

\subsection{Image Reconstruction Based on Quantum Network}
The reconstruction process can be the opposite of the compression process. Specifically, the reconstruction network $U_R$ can be the combination of the quantum gates in the compression network, which are connected in reverse order, so the network parameters need to be retrained. That is because the inverse matrix $U_C^{-1}$ of the compression network $U_C$ can be used directly as the reconstruction network $U_R$ ($U_R=U_C^{-1}$) only when the error of the compressed network is tiny. When the errors are not almost zero, the retrained reconstructed network is more applicable. The output states of the retrained reconstruction network can be represented as:

\begin{align}\label{eq4}
    |\Psi_i\rangle=\sum_{j=0}^{N-1}{B_{i}^{j}}|j\rangle =U_RP_1U_C|\psi_i\rangle=U_RP_1U_C\sum_{j=0}^{N-1}A_i^j|j\rangle,
\end{align}
where $B_{i}^{j}$ is the probability amplitude of the $j$-th computational basis state in the $i$-th output state $|\Psi_i\rangle$ in $U_R$. The probability amplitude of the output state is obtained by measuring the state $|\Psi_i\rangle$, which is then converted into classical information by Eq. (2). Similarly,  the loss between the measurement results and the target is utilized to optimize the parameters of the reconstruction network.

\subsection{Loss Function}
The training target of the quantum compression and reconstruction network consists of two parts: the compression target and the reconstruction target. The goal of the compression network is to compress entangled states into a $d$-dimensional Hilbert space with a high enough attitude fidelity to achieve compression of quantum states. The goal of the reconstruction network is to reconstruct the quantum states of the $d$-dimensional Hilbert space into the $N$-dimensional Hilbert space while making the fidelity of the states too close to 100\%.
Therefore, under given compression and reconstruction targets, complete square variance is used as a loss function, expressed as follows:
\begin{align}\label{eq5}
\begin{cases}
L_C=\sum\nolimits_{j=0}^{N-1}{\sum\nolimits_{i=1}^M{\left( a_{i}^{j}-b_{i}^{j} \right) ^2}}\\
L_R=\sum\nolimits_{j=0}^{N-1}{\sum\nolimits_{i=1}^M{\left( B_{i}^{j}-A_{i}^{j} \right) ^2}},\\
\end{cases}
\end{align}
where $L_C$ and $L_R$ represent the loss functions of quantum compression and reconstruction, respectively. The sample number is $M$, $b_i^j$ represents the certain target probability amplitude of the output state in the compression network, and $A_i^j$ can be directly used as the output target in the reconstruction network. Such as $(b_i)^2=[0,0,0,0,0.25, 0.25, 0.25, 0.25]$ are compression target amplitudes and $(A_i)^2= [0.1,0.1,0.1,0.1,0.1,0.1,0.1,0.1]$ for 8-dimensional data using 3 qubits. Correspondingly, $A_i^j$ needs to be compressed into approximately the 4-dimensional space as $b_i^j$. Since the classical data are all real numbers, the loss function values are also real numbers, and the constructed networks are all real numbers.

\begin{figure*}[htbp]\label{F2}
\centerline{\includegraphics[scale=0.63]{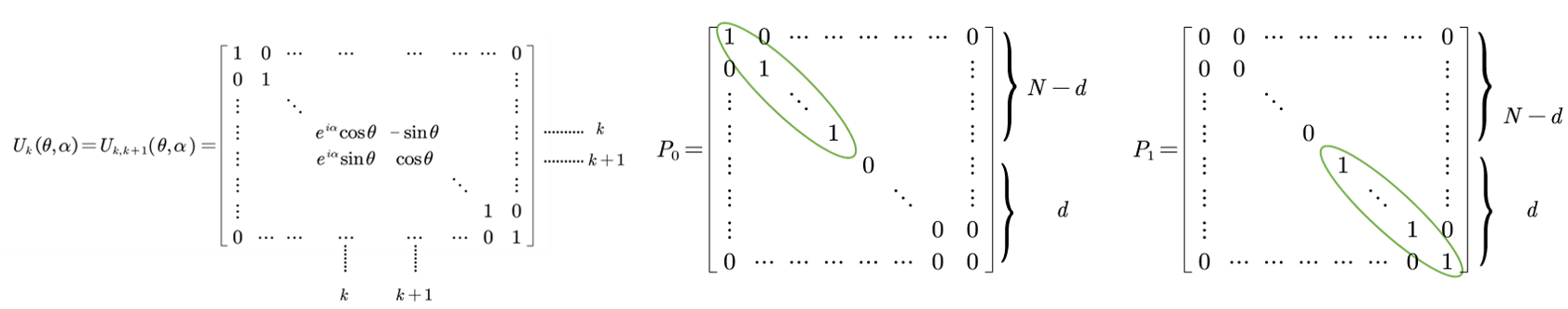}}
\caption{Quantum Gates: $U_k, P_1, P_0$. $U_k$ is quantum gate, $P_1$ and $P_0$ are projection transformation for compression.}
\end{figure*}

\section{Quantum Network}
\subsection{Quantum Gate}
At present, there are many quantum systems and devices that can effectively realize quantum circuits, and the quantum network proposed by us is more suitable for optical quantum circuits. Its circuit structure is flexible, which makes it easy to build complex quantum circuits and realize the scalability of quantum computing.
An ideal lossless multiport optical interferometer is used for transformation between $N$-dimensional vector spaces, which can be described by an $N \times N$ unitary rotation scattering matrix $U_{(k, k+1)}$ \cite{b19}. Equivalently, in quantum optical circuits, $U_{(k, k+1)}$ describes the conversion from the creation or annihilation operator of the input module to the creation or annihilation operator of the output module. In this implementation, the transformation between the $k$-th and $(k+1)$-th vector spaces corresponds to a lossless beam splitter, where the reflectivity and the phase shift are represented as $\cos\theta(\theta \in[0,\pi/ 2])$ and $\alpha(\alpha\in[0,2 \pi])$ respectively at $k$-th input. Then we simplify the symbol $U_{(k, k+1)}$ to $U_k$, which can be described in Fig.2.
Moreover, we always keep the phase shift $\alpha$ constant to 0 ($\alpha\equiv0$), that is, the quantum gate is always a real quantum gate and we just need to train $\theta$.

\subsection{Structure of Quantum network}

The combined quantum gate $U$ represents the complete all-qubit homogeneous transformation change, and $U_{(k, k+1)}$ represents the transformation quantum gate of the $k$-th and $(k+1)$-th vector spaces. Conversion of all qubits requires at least $N-1$ quantum gate combinations. The connection of the quantum rotation gate can be in order. Among them, the continuous transformation consisting of $N-1$ quantum gates is regarded as a one-layer quantum gate combination $U$, while the actual network requires a multi-layer quantum gate combination. The specific structure of $U$ is denoted as follows:
\begin{align}\label{eq6}
    U=U_{\left( 1,2 \right)}U_{\left( 2,3 \right)}U_{\left( 3,4 \right)}\left[ \cdots \right] U_{\left( k,k+1 \right)},
\end{align}
The structure of the network can be explained by giving an illustration: for a schematic diagram of a 2-layer quantum network in 8-dimensional, the components of each layer $U$ is expressed as $U=U_{\left( 1,2 \right)}U_{\left( 2,3 \right)}U_{\left(3,4 \right)}U_{\left( 4,5 \right)}U_{\left( 5,6 \right)}U_{\left( 6,7 \right)}U_{\left( 7,8 \right)}$, which is composed of 7 gates in total as shown in Fig. 3. And the gates in reconstruction network are connected in reverse order of $U$. This structure is commonly implemented by optical quantum circuits in previous studies \cite{b19}.

\begin{figure}[htbp]
\centerline{\includegraphics[scale=0.85]{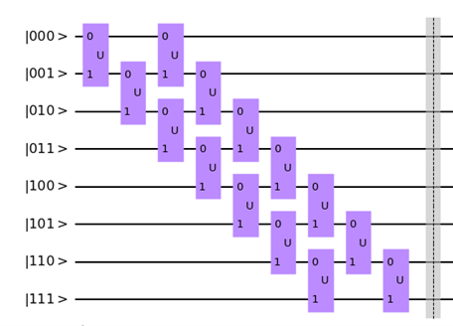}}
\caption{The structure of quantum network. In each layer of this quantum network, the quantum gate $U_{(k, k+1)}$ is connected in Gray-code order. The number of single-layer quantum gates $U$ is $N-1$.}
\end{figure}

\subsection{Quantum Gradient Descent Algorithm}
For quantum networks, the core part is to train the reflectivity parameters $\theta$ of all effective quantum gates. The specific data set is selected for online simulation training, and the actual physical implementation can be set into the corresponding interferometer. The gradient descent (GD) algorithm is used to train and update the parameter $\theta$ \cite{b14} for parameter updates. Next, a gradient descent strategy for quantum compression and reconstruction of network parameters is proposed.
\begin{equation}\label{eq7}
\begin{aligned}
\frac{\partial L_{C}}{\partial \theta _{k}^{p}}&=\frac{\partial \left( \sum\nolimits_{j=0}^{N-1}{\sum\nolimits_{i=1}^M{\left( a_{i}^{j}-b_{i}^{j} \right) ^2}} \right)}{\partial \theta _{k}^{p}}\\
&=2\sum\nolimits_{j=0}^{N-1}{\sum\nolimits_{i=1}^M{\left[ \left( a_{i}^{j}-b_{i}^{j} \right) \frac{\partial \left( a_{i}^{j} \right)}{\partial \theta _{k}^{p}} \right]}}\\
&=2\sum\nolimits_{j=0}^{N-1}{\sum\nolimits_{i=1}^M{\left[ \left( P_1U_CA_{i}^{j}-b_{i}^{j} \right) g_{C}\left( \theta _{k}^{p} \right) \right]}},
\end{aligned} 
\end{equation}
where $\frac{\partial L_{C}}{\partial \theta _{k}^{p}}$ indicates represents the bias (gradient) of the quantum network $L_{C}$ for each phase parameter $\theta$. $k$ and $p$ represent the $\theta$ in the $k$-th gate in the $p$-th layer of the network. The gradient of each $\theta$ in the network is determined by the residual $(a_i^j-b_i^j)$ and the deviation of $a_i^j$ with respect to $\theta$  ($g_{C}\left( \theta _{k}^{p} \right) $). And the gradient of parameters in the reconstruction network is $g_R$. Of course, for batch parameter updates relative to the training set data, the gradient changes of all the training sets need to be summed. $\theta$ can be initialized randomly or uniformly. Different initialization methods will bring different training effects, and subsequent initialization research has also made progress. According to the limit definition of the derivative, the derivative of the parameters in the network can be defined as:
\begin{equation}\label{eq8}
    \begin{aligned}
    g_{C}\left( \theta _{k}^{p} \right) &=\frac{\partial \left( P_1T_{C}A_{i}^{j} \right)}{\partial \theta _{k}^{p}}\\
    &=P_1\underset{\varDelta \rightarrow 0}{\lim}\frac{\left[ T_{C}\left( \theta _{k}^{p}+\varDelta ,0 \right) -T_{C}\left( \theta _{k}^{p},0 \right) \right]}{\varDelta}A_{i}^{j},
    \end{aligned}
\end{equation}
where the phase parameter is $\alpha\equiv0$. The output error $L_C$ of the compression network is calculated by derivation definition for parameter $\theta$. And $\varDelta$ represents the differential step size of $\theta$, uniformly set to $10^{-8}$. When the sample training set is $+\varDelta$, the partial derivative value of the $\theta$ corresponding to loss is calculated. The partial derivative value is calculated using the whole sample set, and we can use the GD algorithm or batch gradient descent algorithm for larger data. This gives an updated expression for the parameter:

\begin{align}\label{eq9}
\mathrm{update}:  \theta _{k}^{p}(t+1)=\theta _{k}^{p}(t)-\eta \cdot \frac{\partial L_{C}}{\partial \theta_{k}^{p}},
\end{align}
where $\eta$ is the learning rate and $t$ is the $t$-th training iteration.

Direct gradient descent (GD) is the most direct and convenient method for phase parameter training. It is suitable for the lightweight quantum gate network parameter training we proposed. According to the gradient descent algorithm, the reflectivity parameter of the quantum sparse coding-decoding network can be trained linearly within the finite iteration period, and the parameter can also be directly set into the corresponding position interferometer for physical implementation.

The pseudo-code of the algorithm for parameter training is shown in Algorithm 1. Firstly, we need to input the image data set $X$ to be compressed and reconstructed, initialize the parameters, update the parameters through the calculation of the loss function and parameter deflection, and finally return the updated results of $\theta$ and process parameters.

 \begin{algorithm}[htbp] 
		\caption{Calculate the loss function and the gradients to update $\theta$.}%
        \label{A2}
		\begin{align*}
			&{\bf Input:}\ l_{C},l_{R},|\psi \rangle, \left( \theta ,0\right) ,Ite,\eta ,d \text { to get }[M, N]\\ %
			&{\bf Output:}\ \left( \theta _{k}^{l_{C}},0\right) ,\left( \theta _{k}^{l_{R}},0 \right) ,g_{C},g_{R},L\\%
			&{\bf Initialize}\ \varDelta=10^{-8}\\ 
            &\left|\hat{\Psi}\right\rangle =P_1T_{C} |\psi \rangle \\
            &\textbf{Update:}\ \left( \theta _{k}^{l_{C}},0\right) \\
			&\textbf{for} \ {i=1:Ite}\\
			&L_R=\left( a_{i}^{j}-b_{i}^{j} \right)\\
            &\textbf{for} \ {l=1:l_{C}}\\
            &\textbf{for} \ {k=1:N}\\
            &\partial \theta _{k}^{l}  =\left[P_1T_{C}\left( \theta _{k}^{l}+\varDelta ,0\right) -P_1 T_{C}\left( \theta _{k}^{l},0 \right) \right] /\varDelta \\
            &g_{C}\left( \theta_{k}^{l} \right)=2sum\left( L_R\cdot \partial \left( \theta _{k}^{l} \right) \right) /\left( M\times N \right)\\
            &\theta _{k}^{l_{C}}=\theta _{k}^{l_{C}}-\eta g_{C}\left( \theta \right)\\
            &{\bf end}\\
            &{\bf end}\\
            &\textbf{Update: }\ \left( \theta _{k}^{l_{R}},0 \right)\\
            &\textbf{the same way to calculate}\ g_{R}\ \\
            &{\bf end}
		\end{align*} 
   \end{algorithm}
\begin{figure*}[htbp]
\centerline{\includegraphics[scale=0.095]{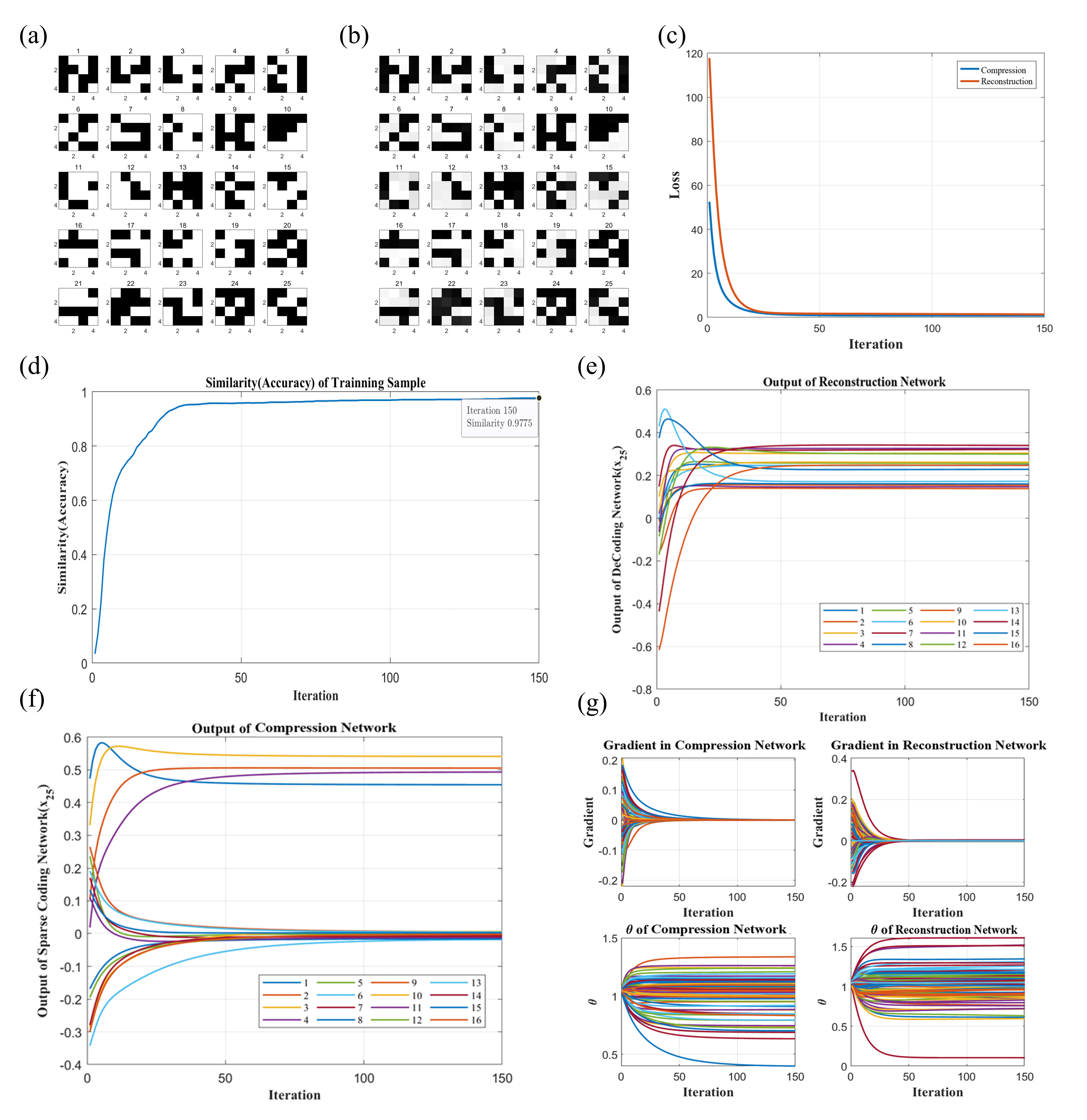}}
\caption{The training process of image compression and reconstruction. \textbf{a} The input binary image in $4\times4$ dimension ($x_i \in \{0,1\},M=25$). These images need to be encoded as the amplitudes of 3-qubit states. \textbf{b} The results of image reconstruction. These images are grayscale ($x_i \in [0,1]$). \textbf{c} The training loss of (\textbf{a}). The $L_C$ and $L_R$ will be almost 0 after 150 training iteration ($minL_C=0.017, minL_R=0.0.023)$. \textbf{d} The reconstruction accuracy of (\textbf{a}). The maximum accuracy is 97.75\%. The compression and reconstruction of (\textbf{a}) $a_{25}$ is (\textbf{f}) and (\textbf{e}) in 150 training iterations respectively. The amplitudes are trained near the target value and stabilize after 50 training iterations. And (\textbf{g}) is the update of $\theta$ in 150 training iterations. The update gradient of $\theta$ decrease to 0 and the $\theta$ stabilize in $[0,2\pi]$.}
\end{figure*}

\section{results}
\subsection{Data Set and Parameter Setting}
 We use Matlab to simulate the quantum algorithm process performed on a classical computer.
 Firstly, we choose 25 binary images and set the basic parameters of the network. Those images are $4 \times 4$-dimensional binary images, which means the input vectors are $16 \times 1$-dimensional column vectors, where the quantum states are in 16-dimensional space for four qubits $(N=16, n=4)$ with 25 samples. Homogeneously, the quantum states are input into the quantum compression network $U_C$ and reconstruction network $U_R$ for training. Set network parameters as follows: the layer of $U_C$ ($l_{C}$) is 12, the layer of $U_R$ ($l_R$) is 14, learning rate $\eta=0.01$, compression channels $d=4$ and train iteration Ite $=150$. Then, the output quantum states in the reconstruction network are measured, the loss function error is calculated, and the quantum gate parameters are adjusted, as shown in III. Obviously, we need to train the parameter $\theta$ with $\alpha$ set to 0. Therefore, only $12 \times 15$ parameters are required to train in the compression network, and $14 \times 15$ parameters are involved in the reconstruction network. Then, the independent training method will be utilized to train the parameters of the quantum network. Finally, the reconstruction implementation is shown in Fig. 4.

\subsection{Image Compression and Reconstruction}
We define accuracy as the similarity between input and output images. If $\left|\hat{x}_i-x_i\right| \leqslant 0.01$, then it can be considered that the pixels in the same positions of two pictures are similar, and the cumulative number of similar pixels $S_p\left(S_p \leqslant D^2\right)$ is increased by 1, for an $D \times D$ picture. Then the accuracy is denoted as:
\begin{align}
    S=\frac{S_p}{D^2} \times 100\%
\end{align}
where $S$ is the similarity between input and output images, also called accuracy.

Then, the binary input (Fig. 4a) transmitted in the quantum network will be modified into grayscale images where the positions of white pixels have no change and merely the values of the white pixels change because of the error of floating number computation in the electronic computer. If we control the output to be binary by comparing the output thresholds, it can be determined that the output amplitude $R_i^j$ will be 0 if it is lower than 0.5; otherwise, it will be 1. Therefore, the output (Fig. 4b) is adjusted by threshold judgment (if $\hat{x}_i \leqslant 0.01, \hat{x}_i=0$; if $\hat{x}_i\geqslant 0.99, \hat{x}_i=1$.).

In the training process, we adjust the network parameters according to the compression and reconstruction effects of the images, including the number of quantum network layers and the learning rate. Finally, the minimum values of $L_C$ and $L_R$ in Fig. 4c are almost 0, which means our training is practical. Furthermore, the reconstruction accuracy of the training reaches 97.75\% (Fig. 4d). The data about compression (Fig. 4f) and reconstruction (Fig. 4e) of Figure 25 in Fig. 4a can converge to the target value quickly. Fig. 4g shows the updating process of $\theta$  where the final parameters are stable, the gradient drops to 0, and the parameter $\theta$ is within $[0,2\pi]$.

\begin{figure}[htbp]
\centerline{\includegraphics[scale=0.28]{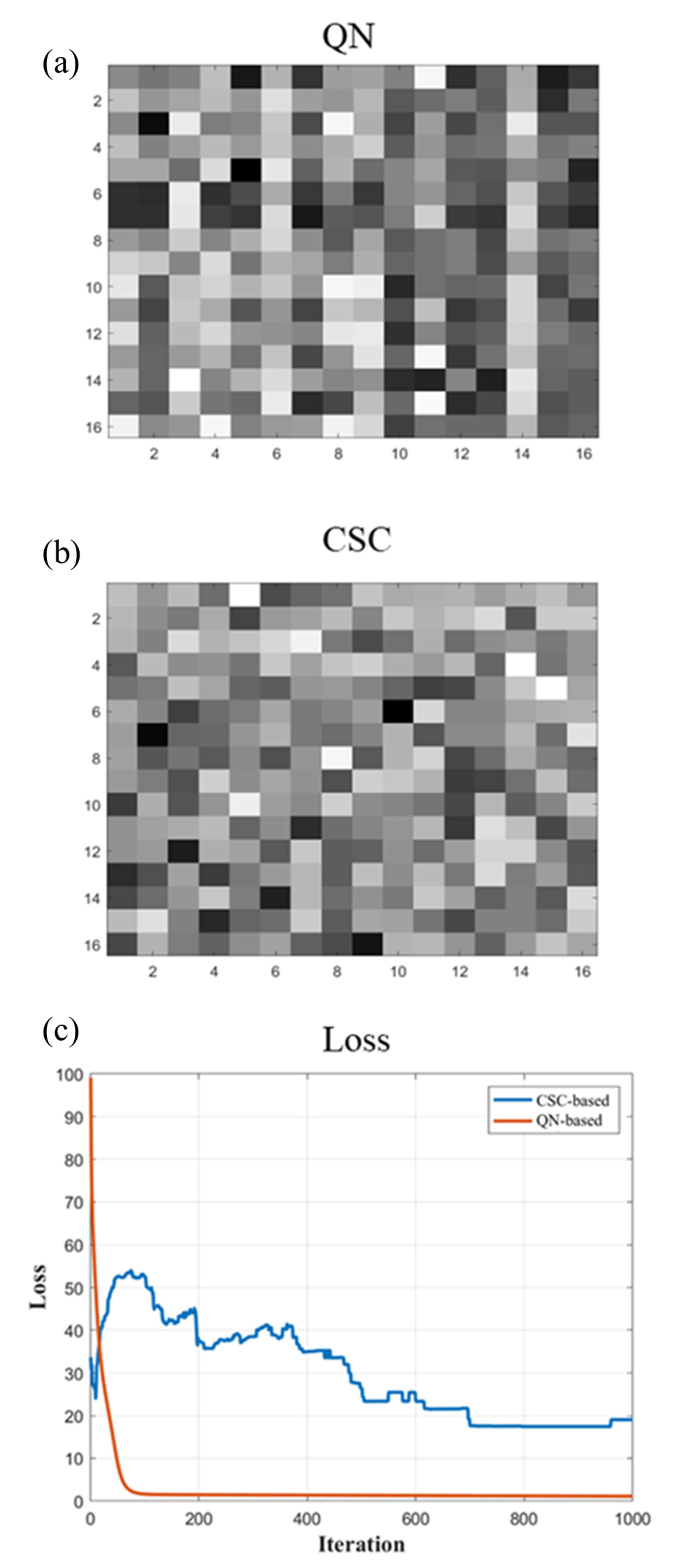}}
\caption{Comparative analysis of QN-based and CSC-based methods. \textbf{a} The quantum compression network $U_C$ in QN-based is $16\times16$ size. \textbf{b} The dictionary set in CSC-based algorithm is $16\times16$ size. \textbf{c} The training loss of compression in the QN-based and CSC-based algorithm.}
\end{figure}

\subsection{Analysis}

Using the same scale network (a $16 \times 16$ dictionary) structure (Fig. 5a and 5b) for the same data set (Fig. 5a), we compared the CSC based on the SVD algorithms \cite{b23} with the QN-based algorithm. Specifically, in the CSC, we can use a sparse coding vector $s$ and a dictionary $D$ to express the input $y$, denoted as $y=D s$ \cite{b23}. We can see the comparison of training losses as shown in Fig. 5. For the same data set, the training time (CPU runs) of the CSC-based algorithm is longer, and the training loss of the QN-based algorithm is much lower than that of the CSC-based algorithm, as shown in Fig. 5c. Therefore, our QN-based algorithm has better quantum superiority than the CSC-based algorithm to a certain extent. Similarly, compared with other quantum algorithms \cite{b11}, the advantages of our proposed quantum network structure can be significantly reflected.

\begin{table}[htbp]
\caption{Quantum Superiority Analysis.}
\begin{center}
\begin{tabular}{|c|c|c|c|}
\hline
\multirow{2}{*}{Method} &\multicolumn{3}{|c|}{\textbf{Index}} \\
\cline{2-4} 
\textbf{}& \textbf{{Accuracy}}& \textbf{{CPU Runs}}& \textbf{{Matrix Size}} \\
\hline
QN-based & 97.75\% & 575.67s & 16*16 \\
\hline
CSC-based & 93.63\% & 763.83s & 16*16   \\
\hline
\multicolumn{4}{l}{$^{\mathrm{a}}$More indexes can also be considered.}
\end{tabular}
\label{tab1}
\end{center}
\end{table}

\section{Discussion}
In this study, we construct a corresponding quantum algorithm based on classical image compression and reconstruction algorithms, using quantum states as image information carriers. Image compression and reconstruction are realized through quantum compression and reconstruction networks. Then, the reconstructed quantum states are decompiled into classical information, and finally, the compression and reconstruction of classical image information are realized.
The simulation results fully prove the effectiveness and superiority of the image compression and reconstruction algorithm based on quantum networks.
The real quantum network where we set $\alpha\equiv0$ still has certain limitations on quantum problems and can only deal with real information compression and reconstruction problems. Therefore, in the future, it is necessary to retain the phase parameter $\alpha$ in the quantum gates and build a fully complex quantum network, which will be more suitable for more diverse quantum problems. Similarly, by constructing complex quantum networks, we expect they could directly solve the problem of compression and recovery of known or unknown quantum states.
In the future, the classical application fields of quantum algorithms will be continuously expanded, and image compression and reconstruction algorithms based on quantum networks will have the potential to be practically used in the field of quantum vision. The QN-based algorithm is a calculation that can be applied to general quantum imaging based on optical quantum circuit design. We will give the implementation and simulation results performed on a quantum computer.

\section*{Acknowledgment}

This work is supported by the National Natural Science Foundation of China (No. 12175104), the National Key Research and Development Program of China (No. 2023YFC2205802), and the Innovation Program for Quantum Science and Technology (No. 2021ZD0301701).

\vspace{12pt}

\end{document}